\documentclass{PoS}
\PoS{PoS(LAT2005)017}
\usepackage{epsfig}
\usepackage{rotating}

\newcommand{\Z}{{\sf Z \!\!\!\! Z}}
\newcommand{\1}{{1 \!\!\! 1}}
\newcommand{\Tr}{\mbox{Tr}}

\title{Confinement and the center of the gauge group}

\ShortTitle{Confinement and the center of the gauge group}

\author{\speaker{Michele Pepe}
\thanks{This work is supported by the Schweizerischer Nationalfonds.}\\

         Institute of Theoretical Physics, University of Bern,
          Sidlerstrasse 5, CH-3012 Bern, Switzerland\\
        E-mail: \email{pepe@itp.unibe.ch}}

      \abstract{The question of the role of the center of the gauge group in the
        phenomenon of confinement in Yang-Mills theory is addressed. The investigation is
        performed from the most general perspective of considering all possible choices
        for the gauge symmetry group. In this context, an interesting role is played by
        $G(2)$ Yang-Mills theory: the simplest pure gauge theory with a trivial center and
        without 't~Hooft flux vortices. Numerical simulations show the presence of a first
        order finite temperature deconfinement phase transition in $G(2)$ Yang-Mills
        theory in (3+1) dimensions. Interestingly, the $G(2)$ gauge symmetry can be broken
        to an $SU(3)$ subgroup by the Higgs mechanism. We investigate the relation between
        the deconfinement phase transition in $G(2)$ and $SU(3)$ Yang-Mills theories by
        numerical simulations in the $G(2)$ gauge-Higgs system.}

\FullConference{XXIIIrd International Symposium on Lattice Field Theory\\

                 25-30 July 2005\\

                 Trinity College, Dublin, Ireland}

\begin{document}
\section{Introduction and Motivation}
What is the role of the center of the gauge group in the deconfinement phase transition of
Yang-Mills theory? Answering this question is a very challenging problem and many
investigations have been carried out in order to address this issue. The center of the
gauge group has turned out to be relevant in two main aspects of confinement. The first
one concerns the critical behaviour at the deconfinement phase transition and the second
one is the potential mechanism of confinement in Yang-Mills theory. We will now examine
these two features. This will allow us to discuss why it is interesting to study how
confinement shows up in a gauge theory with a symmetry group with a trivial center like
$G(2)$.

\subsection{Order of the deconfinement transition: a general approach}\label{order}
In this section we discuss the order of the deconfinement phase transition in Yang-Mills
theory. The approach is general and all possible gauge symmetry groups are taken into
account~\cite{Hol03,Hol03a,Pep04}.

The Yang-Mills theory with gauge group $G$ behaves differently at low and at high
temperatures. The low-temperature regime is characterised by the dynamics of glueballs,
which are color-singlet bound states of gluons. At high temperatures gluons are no longer
confined into glueballs and form a plasma. In general, these two regimes are separated by
a phase transition where the spontaneous breaking of a global symmetry related to the
center ${\cal C}(G)$ of $G$ occurs. The Polyakov loop $\Phi(\vec x)$ corresponds to
inserting a static quark into the gluon system and its expectation value $\langle \Phi
\rangle = \exp (-F/T)$ can be interpreted as a quantity related to the free energy $F$ of
a static quark at temperature $T$. The Polyakov loop is an order parameter for center
symmetry breaking since it has the non-trivial transformation rule $\Phi(\vec x)'= z\;
\Phi(\vec x)$ under the center transformation characterised by the center element $z\in
{\cal C}(G)$. In the low-temperature phase, quarks are confined into hadrons and the free
energy of a quark is infinite: $\langle \Phi \rangle = 0$ and the center symmetry is
unbroken. Instead, in the high-temperature phase, the free energy $F$ has a finite value:
$\langle \Phi \rangle \neq 0$ and the center symmetry is spontaneously broken.

Some time ago Svetitsky and Yaffe~\cite{Sve82} have conjectured a connection between the
critical behaviour of a gauge theory at the deconfinement transition and the critical
behaviour of a scalar theory with a symmetry corresponding to the center of the gauge
group.  Integrating out the spatial degrees of freedom of the $(d+1)$-dimensional
Yang-Mills theory, one can write down a local effective action for $\Phi$. This effective
action describes a scalar field theory in $d$ dimensions with a global symmetry ${\cal
  C}(G)$.  The confined phase of the Yang-Mills theory corresponds to the disordered phase
of the scalar model since the center symmetry is unbroken. The deconfined phase has,
instead, its counterpart in the ordered phase. 

The effective action has, in general, a complicated form and it depends on a number of
parameters that need to be tuned in order to match with the Yang-Mills theory and
reproduce its critical behaviour ~\cite{Dum00}. However, if the deconfinement phase
transition is second order, approaching criticality the correlation length diverges and
the critical behaviour is universal. Hence, the details of the local effective action
become irrelevant: only the center symmetry ${\cal C}(G)$ and the dimensionality $d$ of
space determine the universality class.

Svetitsky and Yaffe's conjecture assigns the center an important role in the critical
dynamics of the Yang-Mills theory at the deconfinement phase transition. Many numerical
simulations have investigated the deconfinement phase transition in Yang-Mills theory and
they have successfully checked the validity of that conjecture in the cases of second
order deconfinement phase transition. These studies have mainly considered the Yang-Mills
theory with gauge group $SU(N)$. However, if one wants to investigate the relevance of the
center in determining the order of the deconfinement phase transition, then the $SU(N)$
groups are not the only and not even the best choice. In fact, since the center of
$SU(N)$ is $\Z(N)$, when changing the group the center changes as well.  This implies that
the symmetry of the effective scalar field theory is modified and the eventually available
universality class changes too. Hence one should widen the perspective and consider other
possible gauge groups. In (\ref{tabgroups}) we list all possible gauge simple Lie groups
with their center subgroups.
\vskip-1.cm
\begin{equation}\label{tabgroups}
\begin{array}{c}
{\cal C}(SU(N))=\Z(N);\hskip.3cm {\cal C}(Sp(N))=\Z(2) ;\hskip.3cm
{\cal C}(SO(N))\hskip-.1cm\rightarrow \hskip-.1cm
{\cal C}(Spin(N))= \hskip-.1cm
\left\{\begin{array}{ll}
\Z(2);   &\;\; N \;\mbox{odd} \\
\Z(2)^2 ;&\;\; N=4k \\
\Z(4) ;  &\;\; N=4k+2
\end{array}\right.\\ 
{\cal C}(G(2))= {\cal C}(F(4))={\cal C}(E(8))=\{\1\};\hskip.4cm
{\cal C}(E(6))=\Z(3);\hskip.4cm {\cal C}(E(7))=\Z(2)
\end{array}
\end{equation}

Besides the $SU(N)$ branch, there are two other ones. The first one is the $Sp(N)$ branch
that is characterised by the property that all its members have the same center $\Z(2)$.
The second branch is that of the $SO(N)$ groups. Contrary to the other ones, these groups
are not simply connected and their universal covering groups are $Spin(N)$: they have
center $\Z(2)$ for $N$ odd and center $\Z(2) \times \Z(2)$ or $\Z(4)$ when $N$ is an even
or an odd multiple of 2, respectively. In this discussion of the deconfinement phase
transition with a general group, we will assume that different groups with the same
algebra have the same continuum limit. In particular, we will consider the universal
covering group associated to a given algebra. Several numerical investigations concerning
this issue and supporting this viewpoint have been carried out~\cite{Gre81}-\cite{Bar03}.
Finally we have 5 other groups which are exceptional: 3 of them, $G(2)$, $F(4)$ and $E(8)$
have a trivial center while the other two, $E(6)$ and $E(7)$, have, respectively, the
center $\Z(3)$ and $\Z(2)$.

The groups $SU(N)$ and $SO(N)$ leave invariant the scalar product of $N$ dimensional
complex and real vectors, respectively. The groups $Sp(N)$ leave the scalar product of $N$
dimensional quaternions invariant. While the 5 exceptional groups leave certain forms
involving octonions invariant. 

Looking at (\ref{tabgroups}), the $Sp(N)$ branch turns out to be particularly interesting.
In fact, contrary to $SU(N)$, one can disentangle the size and the center of the group:
one can increase the size of the group keeping the center $\Z(2)$ fixed. Thus, all $Sp(N)$
Yang-Mills theories in (3+1) dimensions may have a continuous deconfinement phase
transition in the universality class of the 3-dimensional Ising model. Interestingly,
$Sp(N)$ groups are another generalisation of $SU(2)=Sp(1)$ than $SU(N)$. Numerical
simulations of $Sp(2)$ and $Sp(3)$ Yang-Mills theories in (3+1) dimensions show evidence
for first order deconfinement phase transitions~\cite{Hol03}. Based also on the results
available in the literature for $SU(N)$ groups, we conclude that the center does not play
a role in determining the order of the deconfinement phase transition. In other words, the
order of the deconfinement phase transition is a strongly dynamical issue that can not be
decided simply by symmetry arguments. On the other hand, as Svetitsky and Yaffe have
pointed out, if the transition is second order then the center determines the universality
class for the phase transition.

We have conjectured~\cite{Hol03,Hol03a,Pep04} that the size of the group determines the
order of the deconfinement phase transition. At low temperature, we have the confined
phase where the relevant degrees of freedom are glueballs, whose number is essentially
independent on the size of the group.  At high temperature, we have, instead, a gluon
plasma where the number of the relevant degrees of freedom increases proportionally to the
size of the group. If the mismatch of the number of degrees of freedom between the two
phases is large, then we can not move continuously from one phase to the other and a first
order deconfinement phase transition shows up. The natural question that now arises
concerns the finite temperature behaviour of Yang-Mills theories for which the gauge group
has a trivial center, like e.g. $G(2)$~\cite{Hol03b}. In fact, in this case, there is no
center symmetry that can break down and a crossover could simply connect the low- and the
high-temperature regimes. However, according to our conjecture about the relevance of the
size of the group in determining the finite temperature behaviour of Yang-Mills theory, we
expect a first order deconfinement phase transition. In fact, $Sp(2)$ has 10 generators
and $Sp(2)$ Yang-Mills theory has a first order deconfinement transition in (3+1)
dimensions. The group $G(2)$ has 14 generators and so we expect $G(2)$ Yang-Mills theory
to have a first order transition\cite{Pep04,Dum03}.

\subsection{The center and mechanisms of confinement}
In this section we discuss the role of the center in the possible mechanism of confinement
in non-Abelian gauge theories. 

Understanding the mechanism of color confinement stands out as one of the most challenging
open theoretical problems of QCD. Despite the long familiarity with this phenomenon and the
huge amount of phenomenological results, the mechanism that confines quarks inside hadrons
still lacks a full and satisfactory explanation. The most fruitful approach is an
effective description of the phenomenon dating back to the late '70s - early
'80s~\cite{Man74}-\cite{tHo81}, when it was proposed that confinement could be related to
topological objects which, after condensation, disorder the system. The candidate topological
objects are instantons, merons, Abelian monopoles and, in particular, 't~Hooft flux
vortices, which are believed to be the most fundamental ones due to their simple
structure.  Many Monte Carlo simulations have been carried out in order to study these
topological objects~\cite{Kro87a}-\cite{Gre03}. If we consider a Yang-Mills theory with
gauge group $G$, 't~Hooft flux vortices are present if the following inequality holds
\begin{equation}\label{sectors}
\pi_1(G/{\cal{C}}(G)) \neq \{\1\}
\end{equation}

Most of the numerical investigations involve a partial gauge fixing step to detect
topological objects. However, although the gauge-fixed approach gives interesting and
intriguing results, it is important to emphasise that it is not on a completely solid
theoretical ground due to its gauge dependence and the related non-locality. It
should be considered as an approximate method to extract information about the effective
mechanism of confinement. Hence, alternative approaches and different methods to address
the problem of confinement should also be looked for.

For example, the temperature dependence of the 't~Hooft flux vortex free energy has been
investigated. This is a gauge-invariant quantity and the numerical results confirm the
relevance of vortices in the phenomenon of confinement. Another interesting way to
investigate their role is to consider gauge theories without 't~Hooft flux vortices and
study how confinement shows up.

The simplest pure gauge theory with only trivial 't~Hooft flux vortices has the
exceptional group $G(2)$ as gauge symmetry~\cite{Hol03b}.  Numerical studies in this gauge
theory can add further insight in the mechanism of confinement.  An additional bonus that
$G(2)$ provides us is that it has $SU(3)$ as a subgroup. This allows us to move back and
forth between $G(2)$ and $SU(3)$ Yang-Mills theories by exploiting the Higgs mechanism. In
this way we can study how confinement in $G(2)$ Yang-Mills theory is related to the more
familiar case of $SU(3)$ Yang-Mills theory.

\section{$G(2)$ generalities}
\begin{figure}[htb]
\begin{center}
\vskip-.2mm \hskip.8cm  (a) \hskip5.2cm (b)\\ 
\vskip-.2cm
\epsfig{file=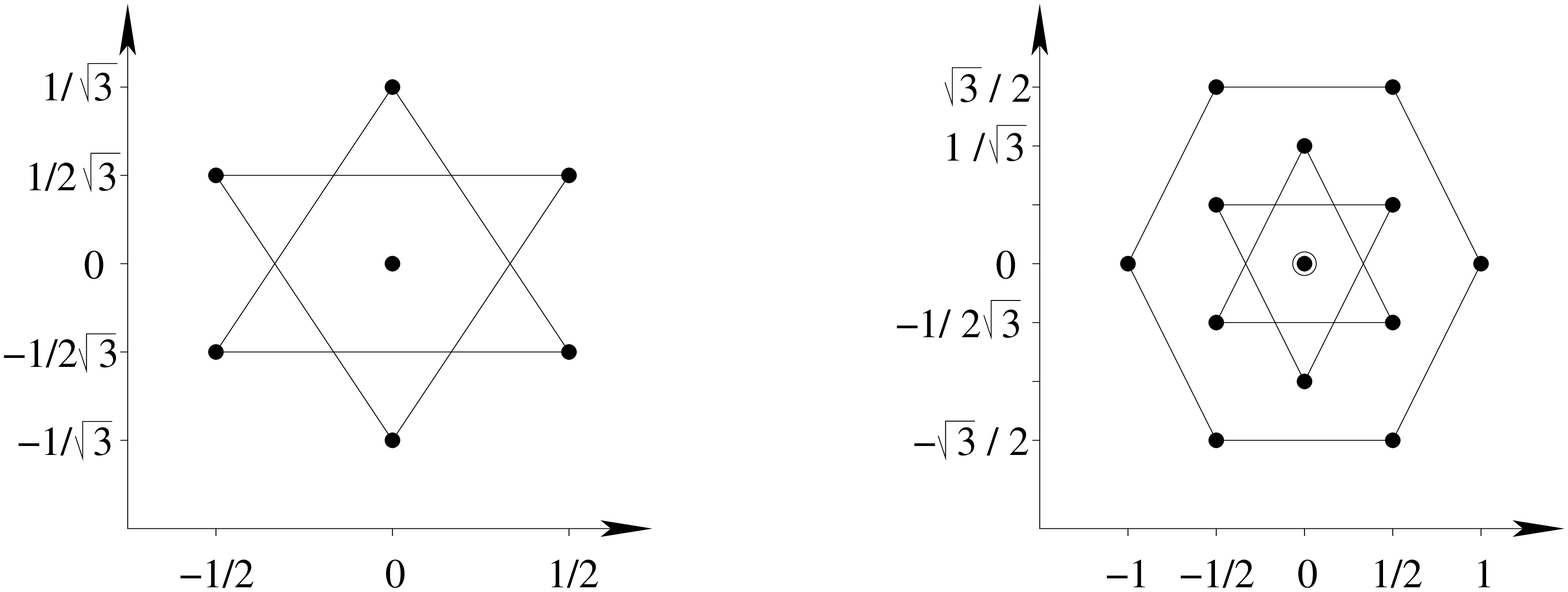,width=10cm}
\caption{Diagrams of the fundamental $\{7\}$ (a) and of the
  adjoint $\{14\}$ (b) representations of $G(2)$.}\label{G2dyag714}
\end{center}
\end{figure}

In this section we present the main features and properties of the exceptional group
$G(2)$.  It is a subgroup of the real group $SO(7)$ which has 21 generators and rank 3.
The $7 \times 7$ real matrices $\Omega$ of $SO(7)$ are defined by the following two
properties
\begin{equation}\label{SO7def}
\det \Omega  = 1; 
\hskip 3cm
\delta_{ab} = \delta_{a'b'} \; \Omega_{aa'}  \;\Omega_{bb'} 
\end{equation}
In addition to these two conditions, the matrices of $G(2)$ satisfy also the constraint
\begin{equation}\label{G2add}
T_{abc} = T_{a'b'c'}\; \Omega_{aa'}\; \Omega_{bb'}\; \Omega_{cc'} 
\end{equation}
where $T_{abc}$ is a completely antisymmetric cubic tensor whose non-vanishing elements -- up to
index permutations -- are 
\begin{equation}\label{Ttensor}
T_{127} = T_{154} = T_{163} = T_{235} = T_{264} = T_{374} = T_{576} = 1
\end{equation}
The additional defining equation (\ref{G2add}) sets 7 constraints on the
$SO(7)$ matrices reducing the initial 21 degrees of freedom to 14, which is the dimension
of $G(2)$. Furthermore the fundamental representation of $G(2)$ is 7 dimensional and so
$G(2)$ inherits from $SO(7)$ the property of having real representations. Hence
$G(2)$-''quarks'' and $G(2)$-''antiquarks'' live in equivalent representations and can not
be distinguished. The rank of $G(2)$ is 2 and so -- similarly to $SU(3)$ -- the diagrams
of its representations can be drawn on a plane. In figure \ref{G2dyag714}a we show the
fundamental representation~$\{7\}$.

The group $SU(3)$ is a subgroup of $G(2)$. When restricting to $SU(3)$, the fundamental
representation $\{7\}$ of $G(2)$ becomes reducible and is given by the sum $\{3\} \oplus
\{\overline{3}\}\oplus \{1\}$. Note that this sum describes a real representation of
$SU(3)$ since $G(2)$ is a real group. This embedding can be made more
explicit in the following basis, where 8 of the 14 $G(2)$ generators can be written as
\begin{equation}\label{su3gen}
\Lambda_a = \frac{1}{\sqrt{2}} \left( 
\begin{array}{ccc} \lambda_a & 0 & 0 \\ 
0 & 0 & 0 \\
0 & 0 & \; -\lambda_a^* 
\end{array} \right).
\end{equation}
Here $\lambda_a$, with $a=1,2,\ldots, 8$, are the usual $3 \times 3$ Gell-Mann matrices. The
$SU(3)$ embedding can also be read off directly from the diagram in figure \ref{G2dyag714}a, where
one can easily identify the triangles of the $\{3\}$, the $\{\overline{3}\}$ and the trivial
representations of $SU(3)$.

The fact that $SU(3)$ is a subgroup of $G(2)$ can be used to derive an explicit
representation of the matrices~\cite{Mac02}. In fact, a $G(2)$ matrix, $\Omega$, can be
expressed as the product of two matrices ${\cal{U}}$ and ${\cal {Z}}$: the former,
${\cal{U}}$, belongs to the $SU(3)$ subgroup while the latter, ${\cal {Z}}$, is in the
coset $G(2)/SU(3) \sim S_6$. The matrices ${\cal{U}}$ and ${\cal {Z}}$
are given by
\begin{equation}
\Omega = {\cal {Z}}(K) {\cal{U}} (U);
\hskip .5cm
{\cal {Z}}(K) = \left( 
\begin{array}{ccc} C(K) & \mu(K)\, K & D(K)^* \\ 
-\mu(K)\, K^\dagger & \frac{1-x(K)}{1+x(K)} & -\mu(K)\, K^T\\ 
D(K) & \mu(K)\, K^* & C(K)^* 
\end{array} \right);
\hskip .5cm
{\cal {U}}(U) = \left( 
\begin{array}{ccc} U & 0 & 0 \\ 
0 &  1 & 0\\ 0 & 0 &\; U^* 
\end{array} \right)
\end{equation}
where $K$ is a $1 \times  3$ complex vector and $U$ is a $3 \times  3$ $SU(3)$ matrix. The
numbers $x(K)$ and $\mu(K)$ are given by $x(K)=||K||^2$ and $\mu(K)=\sqrt{2}/(1+x(K))$,
while the $3 \times  3$ matrices $C(K)$ and $D(K)$ are
\begin{equation}
C(K)= \frac{1}{\Delta}\left\{ \1 -\frac{M}{\Delta (1+\Delta)} \right\};
\hskip .2cm
D(K) = -\frac{W}{\Delta}-\frac{S}{\Delta^2}
\hskip .2cm \mbox{where} \hskip .2cm 
\begin{array}{cc} 
M=K\,K^\dagger; & S = K^*\,K^\dagger \\
W_{\alpha\beta}= \epsilon_{\alpha\beta\gamma} K_\gamma; &
\;\Delta=\sqrt{1+x(K)}
\end{array}
\end{equation}
Using this parametrisation one can see that $\Omega$ depends on 14 real parameters since
the vector $K$ and the $SU(3)$ matrix $U$ depend, respectively, on 6 and 8 real
parameters.

The group $G(2)$ has 14 generators and so the adjoint representation is 14 dimensional:
its diagram is shown in figure \ref{G2dyag714}b. Again, restricting to the $SU(3)$
subgroup, the adjoint representation becomes reducible and it is given by the sum $\{8\}
\oplus \{3\} \oplus \{\overline{3}\}$.  Thus, from the viewpoint of the $SU(3)$ color
degrees of freedom, 8 of the $G(2)$-''gluons'' behave as the 8 gluons and the remaining 6
split into a triplet and an antitriplet behaving like vector quarks and antiquarks.
Similar to the case of the fundamental representation, the reduction of the adjoint
representation of $G(2)$ down to $SU(3)$ can also be read off from the diagram in figure
\ref{G2dyag714}b. Indeed we identify the two triangles corresponding to the $\{3\}$ and the
$\{\overline{3}\}$ and the hexagon of the adjoint representation $\{8\}$ of $SU(3)$.

The fact that both $G(2)$ and $SU(3)$ have rank 2 and that $SU(3)\subset G(2)$, implies
that the center of $G(2)$ is a subgroup of $\Z(3)$, the center of $SU(3)$. The property of
having real representations then makes the center of $G(2)$ trivial. This feature --
together with the property that $G(2)$ is its own universal covering group -- leads to the
absence of non-trivial 't~Hooft sectors:
\begin{equation}
\pi_1 \left( G(2)/{\cal{C}}(G(2))\right) = {\1}
\end{equation}
The triviality of the center also implies that all $G(2)$ representations mix together in
the tensor product decompositions and there is no property similar to triality that
splits the $SU(3)$ representations into 3 branches. Hence, in contrast to $SU(3)$, a
heavy $G(2)$-''quark'' can be screened by $G(2)$-''gluons'' and the string breaks already
in the pure glue theory without the need for dynamical $G(2)$-''quarks''. In particular, from
the tensor product decomposition
\begin{equation}\label{screening}
\{7\} \otimes \{14 \} \otimes \{14 \} \otimes \{14 \} = \{ 1\} + \ldots
\end{equation}
one concludes that 3 $G(2)$-''gluons'' are sufficient to screen a $G(2)$-''quark''.

Finally, as a last feature, we consider the following homopoty groups that tell us which
topological objects can be expected in $G(2)$ Yang-Mills theory
\begin{equation}
\pi_3 (G(2)) = \Z;
\hskip3cm
\pi_2 (G(2)/U(1)^2) = \Z \times \Z
\end{equation}
Hence, there are two types of monopoles just like in the $SU(3)$ Yang-Mills case and
we have instanton solutions leading eventually to non-trivial $\theta$-vacuum effects.

\section{$G(2)$ Yang-Mills theory}\label{G2YM}
In this section we discuss the Yang-Mills theory with gauge symmetry group $G(2)$,
presenting results of numerical simulations at finite temperature.

As we have seen in the previous section, $G(2)$ Yang-Mills theory describes the
interaction of 14 $G(2)$-''gluons''. As any non-Abelian pure gauge theory in 4 dimensions,
it is asymptotically free with a non-perturbatively generated mass-gap. Thus we expect a
confining phase at low temperatures but with a vanishing string tension since the string
between two static $G(2)$-''quarks'' breaks by production of dynamical $G(2)$-''gluons''.
Equation (\ref{screening}) shows that at least 3 $G(2)$-''gluons'' are needed to screen a
$G(2)$-``quark''. As we have seen in the previous section, restricting to the $SU(3)$
subgroup, 8 of the 14 $G(2)$-''gluons'' are related among themselves as the
usual gluons of~QCD~and the remaining 6 split up into a triplet and an antitriplet. Hence,
just like quarks in QCD, these last 6 $G(2)$-''gluons'' explicitly break the center
symmetry $\Z(3)$ of $SU(3)$. Thus, in contrast to~$SU(3)$ Yang-Mills theory, in $G(2)$
Yang-Mills theory we expect the phenomenon of confinement to resemble more closely the one
of QCD but yet without the complications related to dynamical fermions.

Since $G(2)$-''gluons'' screen $G(2)$-''quarks'', the Wilson loop always has -- up to
short distance effects -- a perimeter law and its behaviour can not be considered as an
order parameter for confinement. In other words, the string tension, defined as the
ultimate slope of the static $G(2)$ ``quark''-``quark'' potential, vanishes due to string
breaking and it can not be used to characterise the low-temperature phase of $G(2)$
Yang-Mills theory. One can consider the Fredenhagen-Marcu order parameter~\cite{Fre85}
which is a quantity probing whether the theory has a long range interaction or not, i.e.
if we are in a massless Coulomb phase or in a massive confining/Higgs phase. By a strong
coupling expansion one can show that the theory is indeed in a massive phase.

At finite temperature, due to the triviality of the center of $G(2)$, we expect a
different behaviour than in $SU(3)$ Yang-Mills theory, where confined and deconfined
phases differ by the way the center symmetry is realized. Since in $G(2)$ Yang-Mills
theory no global center symmetry can break down, we can almost exclude a second order
deconfinement phase transition and a crossover can simply connect the low and high
temperatures. However, also a first order transition at finite temperature can take place,
separating the low- and high-temperature regimes. Although a crossover may seem the most
natural possibility, according to the conjecture we have discussed in section \ref{order},
we expect a first order transition. In fact, in the low-temperature regime, the dynamics
is well described in terms of color-singlet glueball states, whose number depends mildly
on the size of the group, i.e. on the number of gluons. On the other hand, at high
temperatures, the large number of gluons, 14, becomes very relevant since we have a gluon
plasma.  Hence the large mismatch in the number of the relevant effective degrees of
freedom between low and high temperatures, suggests a discontinuous behaviour at
finite temperature with a first order phase transition. However the presence of a finite
temperature phase transition and its order are dynamical features that can be properly
addressed only by numerical simulations. In order to investigate these issues we have
performed Monte Carlo simulations of $G(2)$ Yang-Mills theory in 4 dimensions on the
lattice.

\begin{figure}[htb]
\begin{center}
\vskip-.2mm \hskip2cm  (a) \hskip6.7cm (b)\\ 
\hskip-5.2cm
\epsfig{file=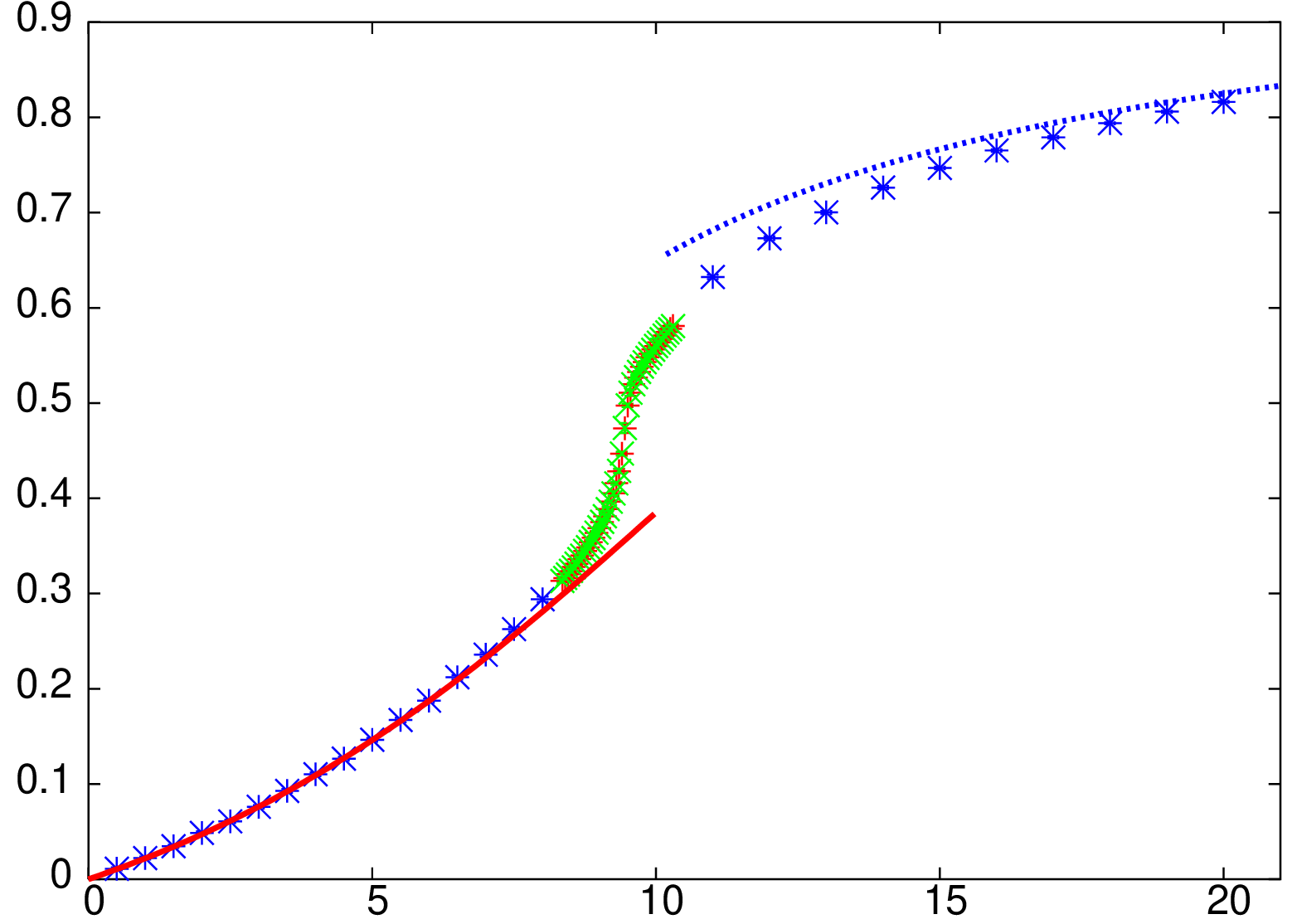,width=6.5cm,angle=0}
\vskip -4.7cm\hskip9cm
\epsfig{file=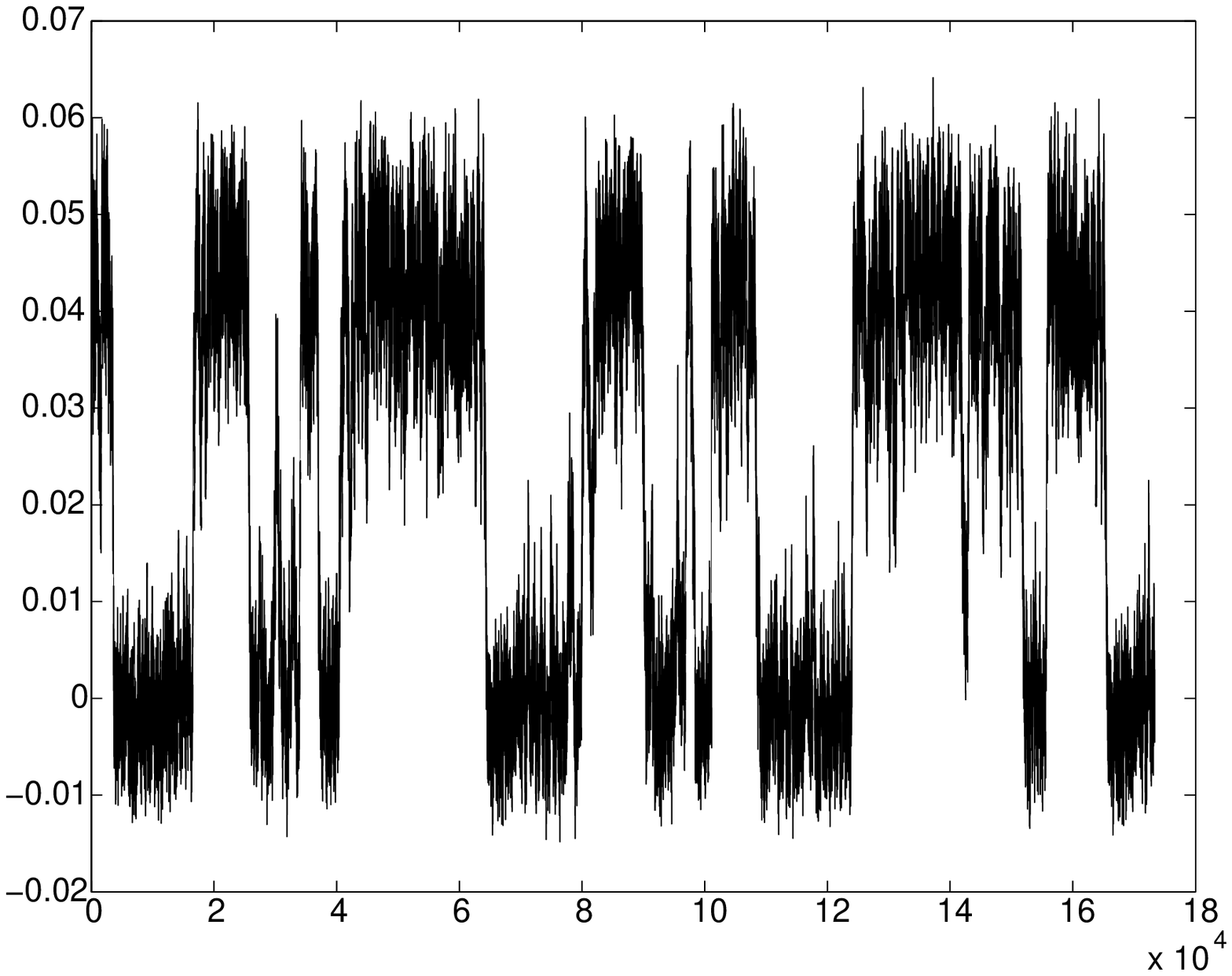,width=5.95cm}
\vskip -3.1cm
\hskip-6cm {\small $\mbox{Tr} \ U_\Box/7$}\hskip7cm {\small $\Phi$}
\vskip 2.2cm
\hskip-.2cm {\small $1/(g^2)$}
\caption{(a): Action density $\mbox{Tr} \ U_\Box/7$ as a function of the gauge coupling
  $1/(g^2)$. (b): Polyakov loop Monte Carlo history from a numerical simulation on a
  $20^3\times 6$ lattice at $1/(g^2)=9.765$.}\label{bulkscan}
\end{center}
\end{figure}

The $G(2)$ Yang-Mills theory on the lattice is constructed in the usual way in terms of
link matrices, $U_{x,\mu}$, which are elements of the group $G(2)$ in the fundamental
representation $\{7\}$ . We consider the Wilson action, which is given by
\begin{equation}
S_{YM}[U] = - \frac{1}{7 g^2} \sum_\Box \mbox{Tr} \ U_\Box =
- \frac{1}{7 g^2} \sum_{x,\mu < \nu} \mbox{Tr} \
\left[
U_{x,\mu} U_{x+\hat\mu,\nu} U^\dagger_{x+\hat\nu,\mu} U^\dagger_{x,\nu} \right]
\end{equation}
where $g$ is the bare gauge coupling. 
As a first step, we have measured the action density at many values of the gauge coupling
in order to check for the presence of a possible bulk phase transition separating the
strong coupling from the weak coupling regime, where the continuum limit can be
approached. In figure \ref{bulkscan}a we show the numerical results. The red and green
points correspond to a fine-grained hysteresis cycle in the region where a bulk transition
could occur. This approach has been used in order to enhance the efficiency in detecting
coexisting phases in case of a bulk transition. Our results indicate that no bulk
transition separates the strong and the weak coupling regimes. The red and the blue lines
are, respectively, the analytic perturbative expansions at strong and weak coupling.
\vskip-5mm

Since no bulk phase transition interfering with the deconfinement phase transition has
been found, we have looked for the presence of a transition at finite temperature.
Consistent with our expectation and contrary to the argument based on the absence of a
center symmetry, we have found that the $G(2)$ Yang-Mills theory has a deconfinement phase
transition at finite temperature~\cite{Pep04}. In figure \ref{bulkscan}b we show the Monte
Carlo history of the Polyakov loop in the critical region. A first order deconfinement
phase transition can be clearly observed, with many tunneling events between two
coexisting phases. It is important to note that, since no symmetry gets broken, no order
parameter can be defined. However discontinuities in physical quantities, for instance in
the free energy of a static $G(2)$-''quark'' or in the specific heat, can be used to
unambiguously detect this finite temperature phase transition. A detailed analysis
requires a finite size scaling study and the continuum limit extrapolation. The value
of the lattice temporal extent, $N_t=6$, and the set of gauge couplings considered in this
study have been chosen in order to be close to the continuum~limit.
\begin{figure}[htb]
\begin{center}
\epsfig{file=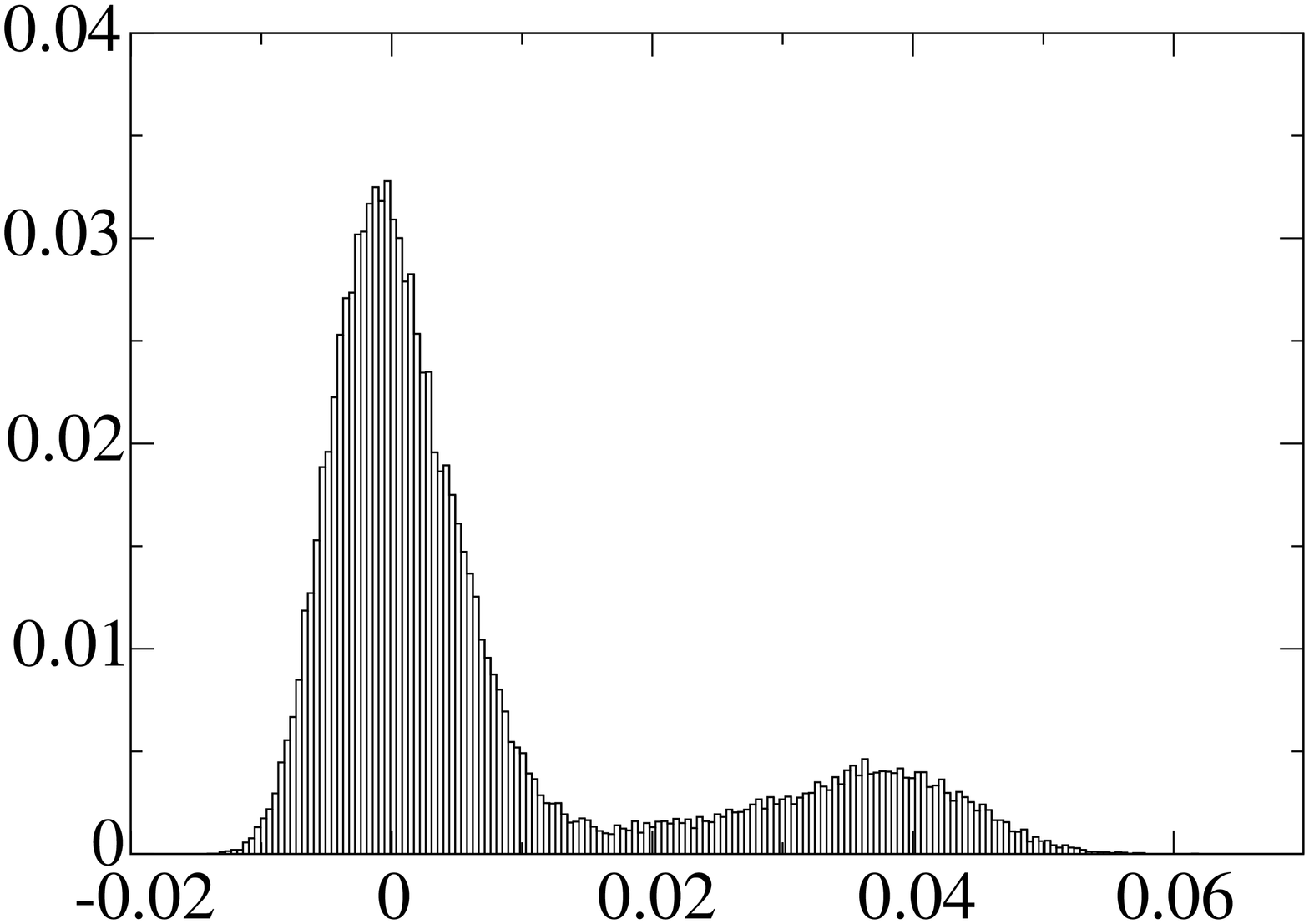,width=3.5cm,angle=-90}
\epsfig{file=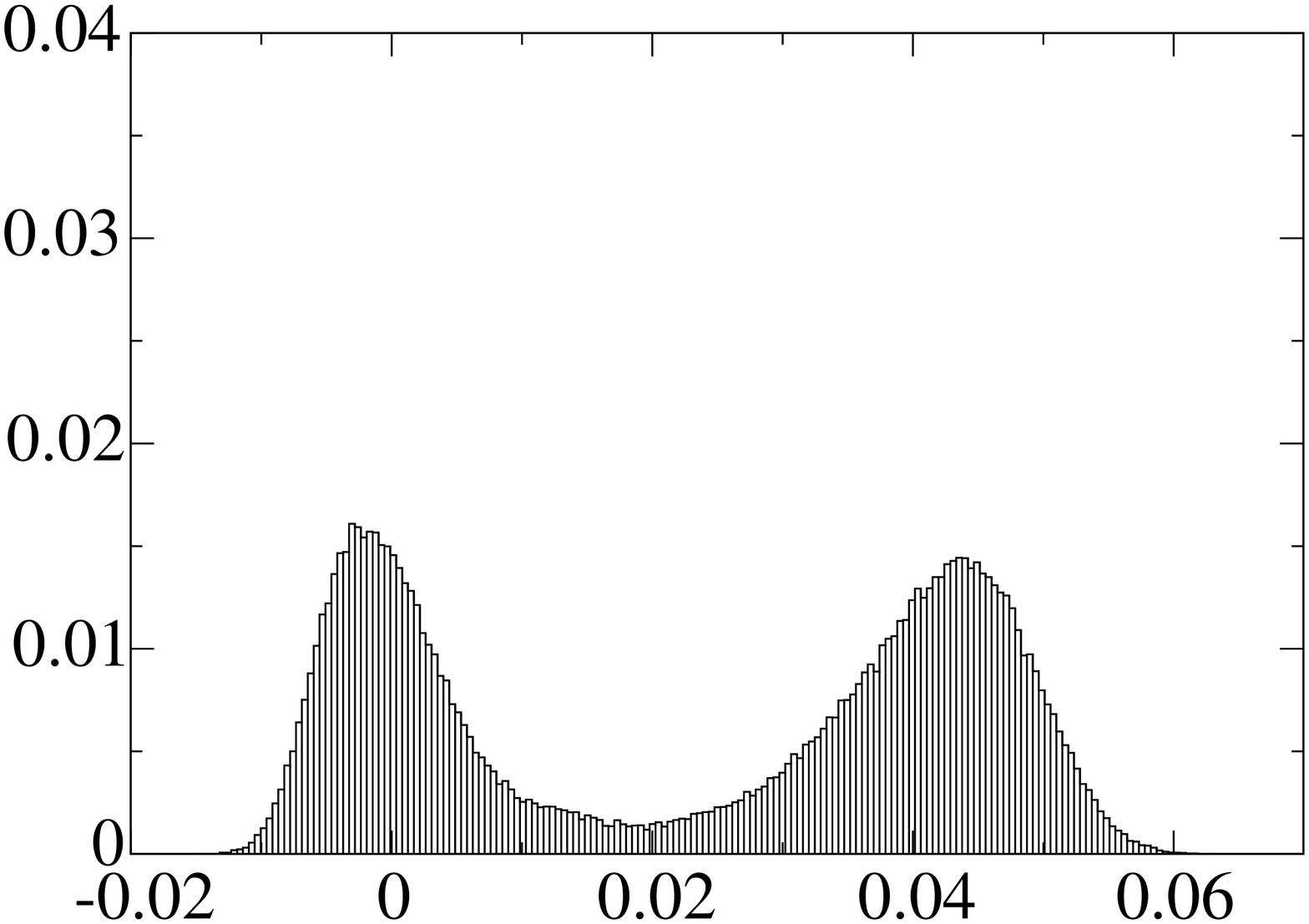,width=3.5cm,angle=-90}
\epsfig{file=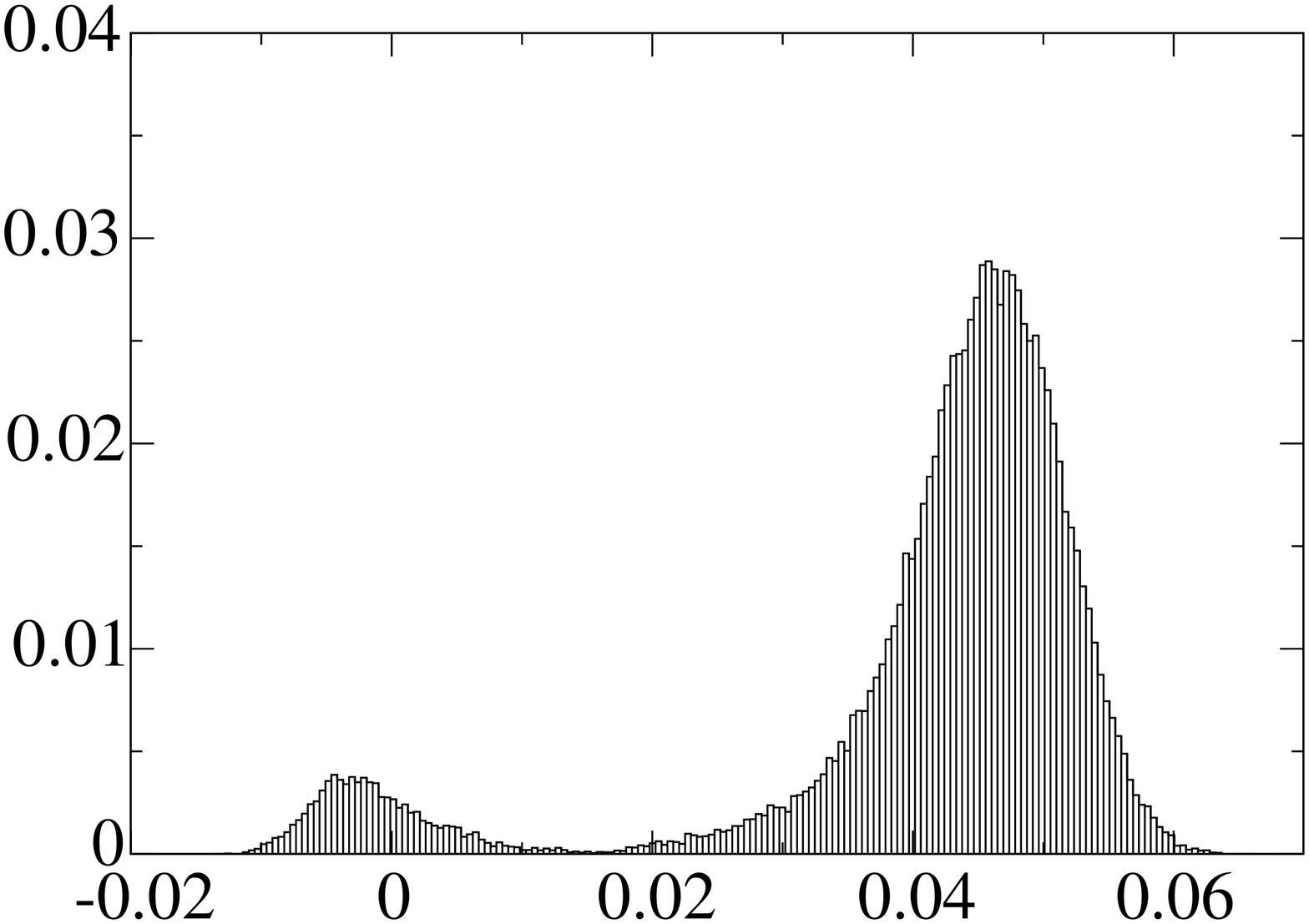,width=3.5cm,angle=-90}
\caption{Polyakov loop probability distributions in the critical region of the
  deconfinement phase transition in the (3+1)-dimensional $G(2)$ Yang-Mills theory. The
  temperature increases from left to right.}\label{critG2YM}
\end{center}
\end{figure}

In figure \ref{critG2YM} we show the probability distributions of the Polyakov loop in the
critical region: the temperature increases from left to right and the switching from one
phase to the other can be clearly observed. Notice that the Polyakov loop in the
low-temperature, confined phase is peaked around zero. However, due to string breaking, we
know that the free energy of a static $G(2)$-''quark'' is always finite. Hence, the very
small value we observe here -- hardly distinguishable from zero -- is related to the fact
that the energy necessary to pop out the $G(2)$-''gluons'' that screen the
$G(2)$-''quark'' from the vacuum is quite large.

\section{The high-temperature effective potential}
In the previous section we have presented the results of numerical simulations in $G(2)$
Yang-Mills theory, showing the presence of a finite temperature, first order deconfinement
phase transition. In order to have a better understanding of the deconfined phase, we
discuss here the analytic perturbative computation of the effective potential for the
Polyakov loop at high temperature.

Perturbative computations at zero and finite temperature differ in an important aspect. At
zero temperature, the temporal component of the gauge field can be eliminated by a gauge
transformation defined up to static gauge transformations. At finite temperature,
instead, the temporal component can not, in general, be gauged to zero but only to a
static value due to the compactified temporal direction. Hence, perturbative expansions
around background gauge fields with different static temporal components are physically
different at finite temperature~\cite{Wei80,Wei81}.

\begin{figure}[htb]
\begin{center}
\vskip-.4mm \hskip.5cm  (a) \hskip7.cm (b)\\ 
\epsfig{file=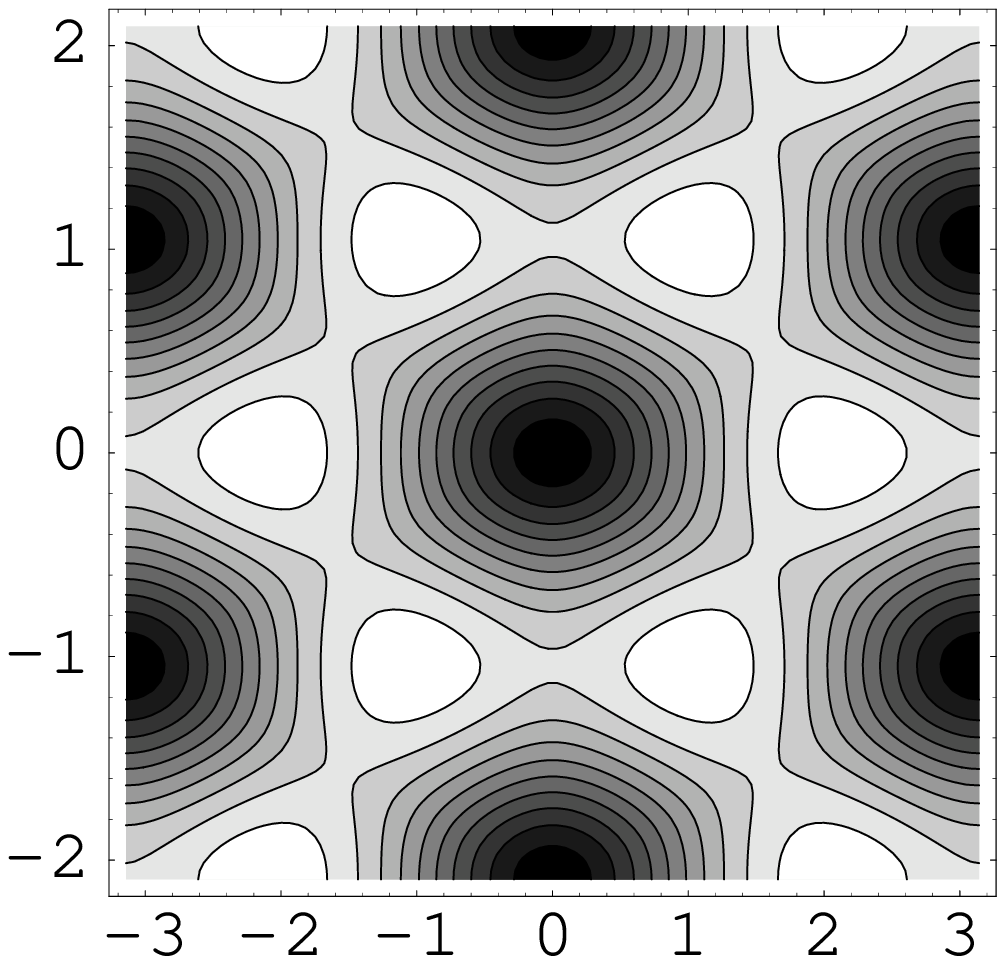,width=4.6cm}\hskip3cm
\epsfig{file=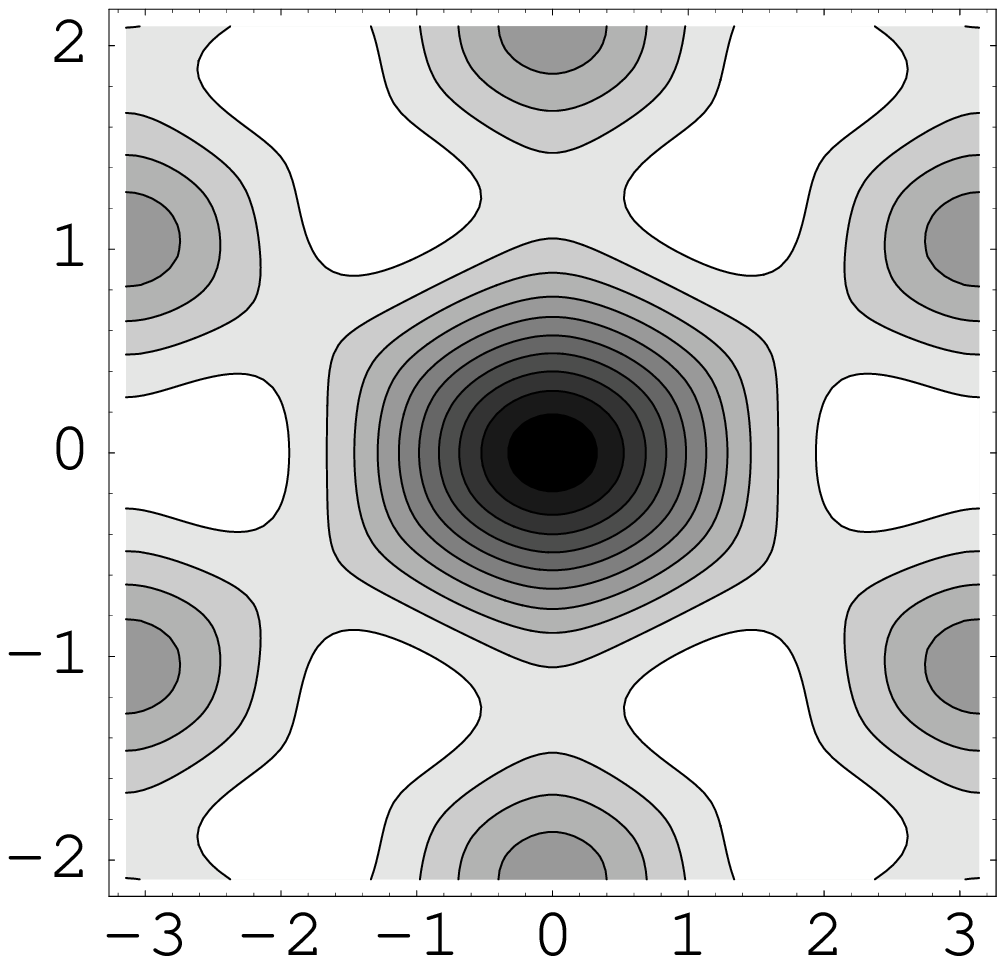,width=4.5cm}
\vskip -2.9cm
\hskip-5cm {\small $\tilde\theta_2$}\hskip7.4cm {\small $\tilde\theta_2$}
\vskip 2.1cm
\hskip.65cm {\small $\tilde\theta_1$}\hskip7.3cm {\small $\tilde\theta_1$}
\caption{Contour plots of the one-loop high-temperature effective potential for the
  Polyakov loop for $SU(3)$ (a) and $G(2)$ (b) Yang-Mills theories.}\label{effpotG2SU3}
\end{center}
\end{figure}

Here we compute the one-loop free energy in the continuum. We perturbe around a general
static background gauge field $A_\mu^B $ given by
\begin{equation}
A_\mu^B  = \delta_{\mu 0} (\sqrt{2}\; \theta_1 \Lambda_3 + \sqrt{6}\;\theta_2 \Lambda_8)
\end{equation}
where $\theta_1$ and $\theta_2$ are the two phases characterising an Abelian $G(2)$
matrix.  Note that evaluating the effective potential for the background field $A_\mu^B$
is equivalent to computing the effective potential for the Polyakov loop since it is
simply the time-ordered integral of $A_\mu^B$ along the temporal direction.  We now
introduce a gauge fixing condition defined by $D_\mu^B A_\mu = 0$, where the covariant
derivative is given by $D_\mu^{ab} = \partial_\mu\delta^{ab} + g f^{abc} A_\mu^c$, with
$f^{abc}$ being the $G(2)$ structure constant. After integrating in the ghost field
$\chi$, the continuum Lagrangian takes the form
\begin{equation}
{\cal {L}} = \frac{1}{4} G_{\mu\nu}^2 +\frac{1}{2\alpha} (D_\mu^B A_\mu )^2 +
\overline{\chi} D_\mu^B D_\mu \chi
\end{equation}
where $\alpha$ is a gauge fixing parameter and the field strength is defined as usual
$G_{\mu\nu}^a = \partial_\mu A_{\nu}^a - \partial_\nu A_{\mu}^a -g f^{abc} A_\mu^b
A_\nu^b$.  We now decompose the gauge field in the sum of the background field and of the
quantum fluctuations $A_\mu^q $: $A_\mu  = A_\mu^B  + A_\mu^q $ . Retaining in
the Lagrangian only terms that are at most quadratic in $A_\mu^q $, the partition
function can be computed and the free energy W is then given~by
\begin{equation}
W[\theta_1,\theta_2] = \frac{1}{2} \Tr \left\{ \log \left[  
\left( \delta_{\mu\nu} (-D^B)^2 \right)+(1-\frac{1}{\alpha}) D_\mu^B D_\nu^B
\right] \right\} 
-\Tr \left[ \log(-(D^B)^2)\right]
\end{equation}
Using some formulas~\cite{Bel89} to manipulate the previous expression, the one-loop free
energy can be explicitly calculated and the result is
\begin{equation}\label{effpot}
W[\theta_1,\theta_2] = \frac{4\pi^2 T^4}{3} 
\left[
-\frac{1}{30} +\sum_{i=1}^6 B_4\left( \frac{C_i(\tilde\theta_1,\tilde\theta_2)}{2\pi}\right)
\right]
\end{equation}
where $B_4(x)=-1/30 +x^2 (x-1)^2$ is the $4^{th}$ Bernoulli polynomial with the argument
defined modulo 1, $1/T$ is the length of the compactified temporal direction and
$\tilde\theta_1=g \theta_1/T$ and $\tilde\theta_2=g \theta_2 /T$. The
$C_i$'s are given by
\begin{equation}
\begin{array}{lll}
C_1(\tilde\theta_1,\tilde\theta_2)= 2 \tilde\theta_1; \;\;\;\; &
C_2(\tilde\theta_1,\tilde\theta_2)=\tilde\theta_1 + 3 \tilde\theta_2; \;\;\;\;  &
C_3(\tilde\theta_1,\tilde\theta_2)=\tilde\theta_1 - 3 \tilde\theta_2; \;\;\;\;\\
C_4(\tilde\theta_1,\tilde\theta_2)= 2 \tilde\theta_2; &
C_5(\tilde\theta_1,\tilde\theta_2)= \tilde\theta_1 - \tilde\theta_2; &
C_6(\tilde\theta_1,\tilde\theta_2)= \tilde\theta_1 + \tilde\theta_2;
\end{array}
\end{equation}
Close to the trivial background field $(\tilde\theta_1,\tilde\theta_2)=(0,0)$, we have 
$W=-(14/45) \pi^2 T^4$, which corresponds to an ideal gas of $G(2)$-''gluons''; note the
number 14 in the numerator corresponding to the number of $G(2)$-''gluons''. Finally, we point
out that, since $SU(3)$ is a subgroup of $G(2)$ with the same rank, the high-temperature
effective potential in $SU(3)$ Yang-Mills theory~\cite{Wei81,Bel89} can be immediately
obtained from (\ref{effpot}) performing the sum only up to 3. In figures
\ref{effpotG2SU3}a and \ref{effpotG2SU3}b we show the contour plots of the one-loop
effective potential for the Polyakov loop at high temperature in $SU(3)$ and $G(2)$
Yang-Mills theories, respectively.

The formula (\ref{effpot}) is, somehow, a counter-intuitive result. In fact, despite the
fact that the center of $G(2)$ is trivial, the center $\Z(3)$ of the $SU(3)$ subgroup
plays a role in the effective potential of $G(2)$ at high temperature (remember that
${\cal{C}}(G(2)) \subseteq {\cal{C}}(SU(3))$). Indeed we have an absolute minimum
corresponding to the trivial center element $(\tilde\theta_1,\tilde\theta_2)=(0,0)$ but we
also have local minima corresponding to the non-trivial $\Z(3)$ center elements
$(\tilde\theta_1,\tilde\theta_2)=(0,\pm 2\pi/3)$ and $(\tilde\theta_1,\tilde\theta_2)=(\pm
\pi,\pm \pi/3)$. The non-trivial $\Z(3)$ center elements are metastable minima since they
are separated from the absolute minimum by a free energy gap that grows proportionally to
the volume: hence they are not relevant in the thermodynamic limit and in determining the
bulk behaviour of the $G(2)$ Yang-Mills theory.  However, numerical simulations are
carried out on lattices of finite size and so there is a finite probability of tunneling
to one of these metastable minima. Then, depending on the algorithm used to update the
system, these metastable states can have long lifetimes and significantly bias the
numerical results.

\section{$G(2)$ Yang-Mills theory with a Higgs field}
In this section we discuss the connection between confinement in $G(2)$ Yang-Mills theory
and in the familiar case of $SU(3)$ Yang-Mills theory. In fact, by the Higgs mechanism,
we can break~the $G(2)$ gauge symmetry down to $SU(3)$ and study how the well-known
form of confinement in $SU(3)$ Yang-Mills theory reemerges as the breaking of the $G(2)$
symmetry becomes stronger and stronger.

The gauge symmetry breaking of $G(2)$ Yang-Mills theory down to $SU(3)$ can be
accomplished by adding a Higgs field in the fundamental representation $\{7\}$ of $G(2)$.
Restricting to the $SU(3)$ subgroup, the adjoint representation $\{14\}$ is reducible and
splits into the sum $\{8\}\oplus\{3\}\oplus\{\overline{3}\}$. Hence, the 8 $G(2)$-''gluons''
which are related among themselves like the gluons, stay massless while the other 6 pick up
a mass $M_G$ proportional to the expectation value $v$ of the Higgs field. Changing $v$, we
can then change the mass of the 6 massive $G(2)$-''gluons''. If $M_G$ is not too large
compared to $\Lambda_{QCD}$, they participate in the dynamics. However, as $M_G$ 
becomes larger and larger, they progressively decouple and, at the end, they are
removed from the dynamics and we are left with the $SU(3)$ Yang-Mills theory.
Thus a Higgs field in the fundamental representation $\{7\}$ provides us with a handle we
can use to interpolate between $G(2)$ and $SU(3)$ Yang-Mills theories.

In $SU(3)$ Yang-Mills theory the behaviour of the Wilson loop is an order parameter for
confinement: the string tension is non-zero in the confined phase and vanishes in the
deconfined phase. On the contrary, in $G(2)$ Yang-Mills theory, the string tension is
always zero. We now discuss how the Higgs mechanism relates these two scenarios. Let us
first consider the pure glue case. As we have discussed in section \ref{G2YM}, the
breaking of the string between two static $G(2)$-''quarks'' happens due to the production
of 3 pairs of $G(2)$-''gluons''.  Hence, the string breaking scale is related to the
dynamical mass of the 6 $G(2)$-''gluons'' popping out of the vacuum. When we switch on the
interaction with the Higgs field, 6 of the $G(2)$-''gluons'' start to pick up a mass for
the Higgs mechanism.  Note now that the two $G(2)$-``quark''/''gluon'' bound states
resulting from string breaking, must be both $G(2)$-singlets and $SU(3)$-singlets. Then,
since the fundamental representation $\{7\}$ reduces to the sum
$\{3\}\oplus\{\overline{3}\}\oplus\{1\}$ when restricting to the $SU(3)$ subgroup, it
follows that, among the 6 $G(2)$-''gluons'' breaking the string, there must be a pair of
the massive ones. Thus the string breaking scale depends on $M_G$ and the larger
$M_G$ goes, the larger is the distance where string breaking occurs. The picture of the
unbreakable $SU(3)$ string then reemerges. When the expectation value of the Higgs
field is sent to infinity, so that the 6 massive $G(2)$-''gluons'' are completely removed
from the dynamics, also the string breaking scale is infinite. Hence, we fully recover the
familiar $SU(3)$ linearly rising, confining potential with a non-vanishing value for the
string tension.

\begin{figure}[htb]
\begin{center}
\epsfig{file=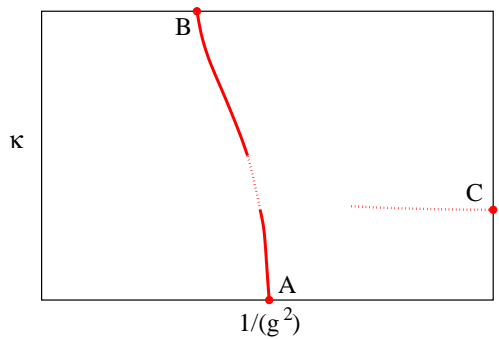,width=6.5cm}
\caption{Phase diagram in the parameter space $(1/(g^2),\kappa)$. The $\kappa=0$ and
  $\kappa=\infty$ axes correspond to the $G(2)$ and $SU(3)$ Yang-Mills theories
  respectively. The $1/(g^2)=\infty$ limit is the $SO(7)$ spin model.}\label{phasediag}
\end{center}
\end{figure}

In order to have a more quantitative understanding of the relation between confinement in
$G(2)$ and $SU(3)$ Yang-Mills theories, we have performed numerical simulations in $G(2)$
Yang-Mills theory with a Higgs field in the fundamental representation $\{7\}$. In
particular, the interesting aspect we want to address is the connection between the
finite temperature critical points in the two pure gauge theories. We consider a Higgs
field $\varphi$ of unit length; the lattice action looks as follows
\begin{equation}
S_{HYM}[U,\varphi]=S_{YM}[U]-\kappa \sum_{x,\mu} 
\varphi_x^\dagger U_{x,\mu} \varphi_{x+\hat\mu}
\end{equation}
where $\kappa$ is the parameter characterising the interaction between the Higgs field and
the $G(2)$-''gluons''. In our numerical study we have considered the same temporal extent,
$N_t=6$, that we have used in the Monte Carlo simulations of $G(2)$ Yang-Mills theory. The
aim of our numerical simulations has been to understand the phase diagram in the
parameter space $(1/(g^2),\kappa)$. The result of this study is summarised in figure
\ref{phasediag}.
\vskip-.5cm

Let us first discuss some particular points in the phase diagram. For $\kappa=0$, the
Higgs field decouples and we have the pure $G(2)$ Yang-Mills theory: the point A in figure
\ref{phasediag} is the corresponding critical coupling of the deconfinement transition.
For $\kappa=\infty$, due to the Higgs mechanism, the 6 massive $G(2)$-''gluons'' are
removed from the dynamics and the $G(2)$ gauge symmetry is completely broken down to
$SU(3)$. The point B in the phase diagram is the critical coupling of the $SU(3)$
Yang-Mills theory at $N_t=6$ (note that a rescaling factor 7/6 has been taken into account
for the different normalisation of the coupling). For $1/(g^2)=\infty$, the gauge degrees
of freedom are completely frozen and we have a spin model with global $SO(7)$ symmetry.
This model has an order/disorder phase transition that is denoted by the point C in the
phase diagram.

\begin{figure}[htb]
\begin{center}
\vskip-2.mm \hskip.05cm  (a) \hskip4.cm (b) \hskip4.cm (c)\\ 
\epsfig{file=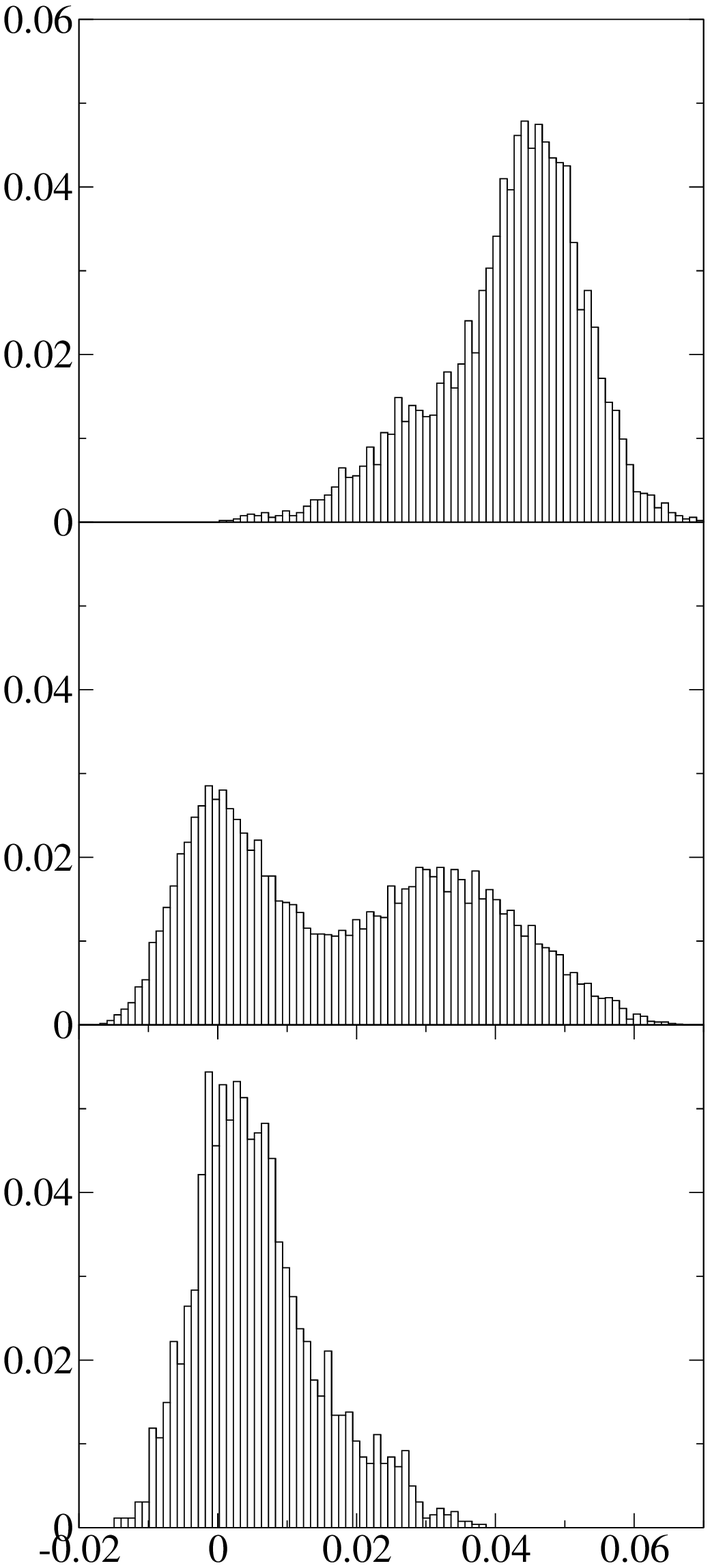,width=3.5cm,height=7.5cm}\hskip1cm
\epsfig{file=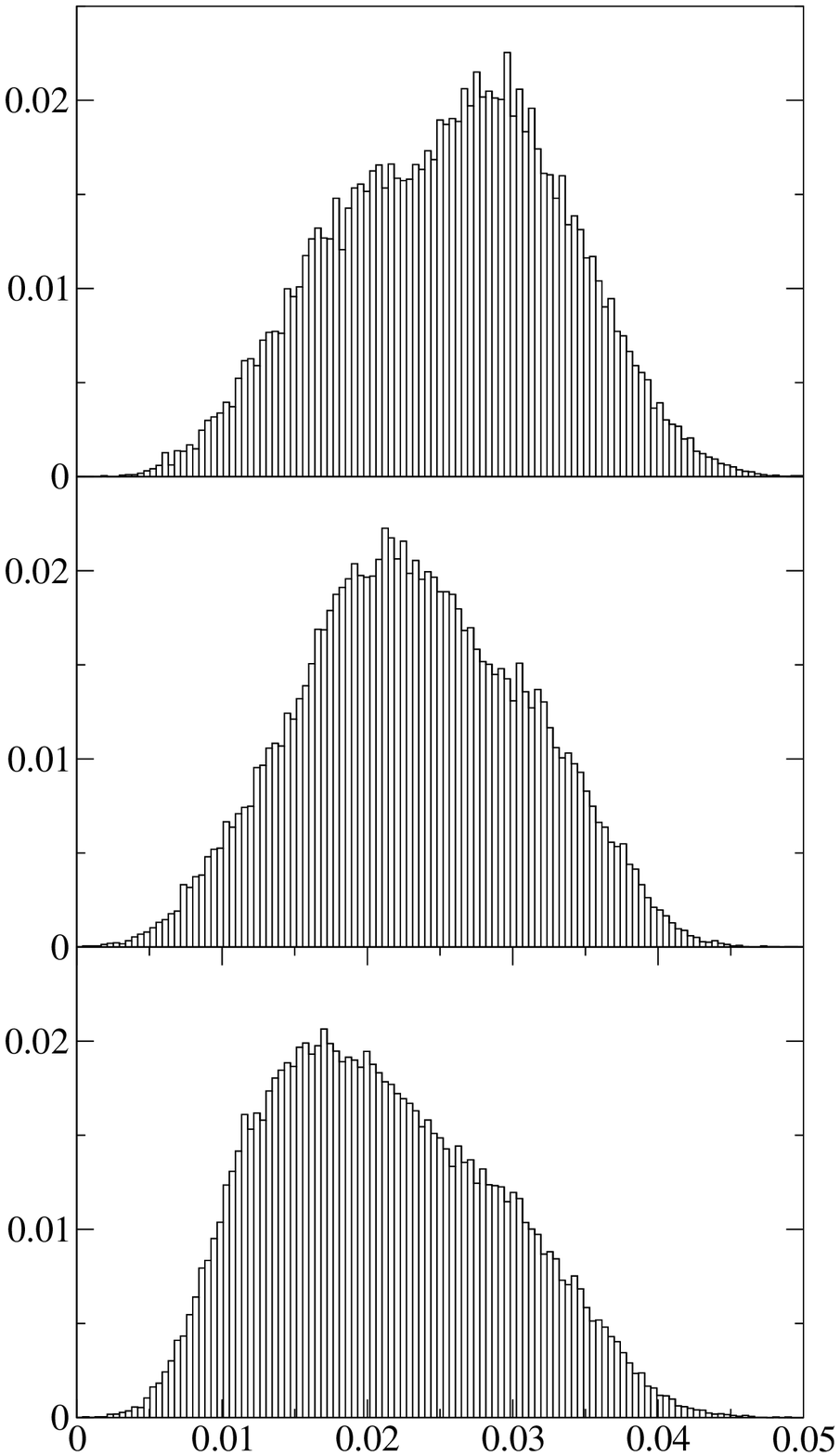,width=3.5cm,height=7.5cm}\hskip1cm
\epsfig{file=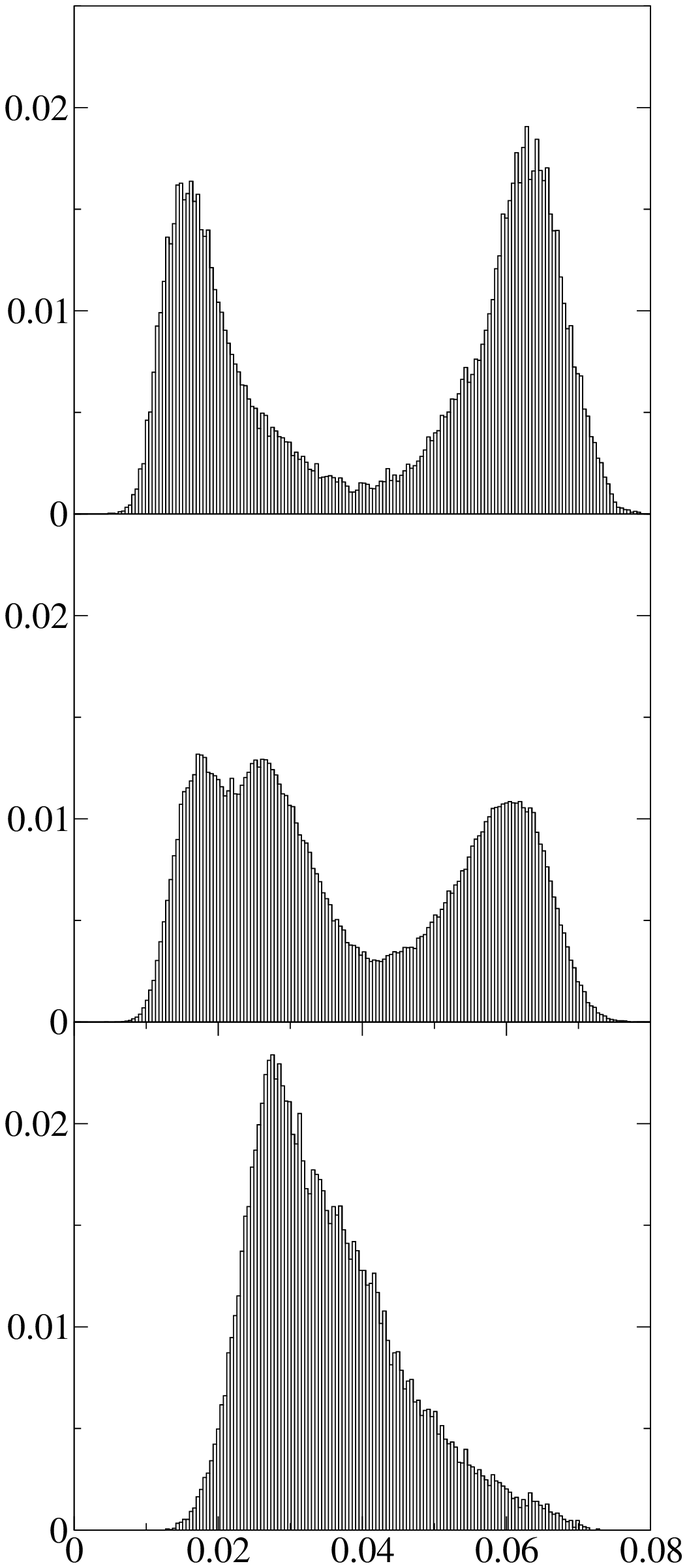,width=3.5cm,height=7.5cm}
\caption{Polyakov loop probability distributions at $\kappa=1.3$ (a), $\kappa=1.5$ (b) and
  $\kappa=4.0$ (c). The gauge coupling $1/(g^2)$ increases from bottom to
  top.}\label{critG2HYM}
\end{center}
\end{figure}

Starting from A, B and C, we have carried out numerical simulations in order to
investigate how these points are related. In particular, our main interest is to study the
relation between A and B. However let us first discuss the point C. We have not found any
indication of a critical line ending at the point C. Nevertheless, since it was not our
main interest studying this part of the phase diagram, we can not rule out a second order
line or a weak first order one that could, eventually, show up considering larger
lattices.  For this reason in figure \ref{phasediag} we have plotted a dotted line ending
at the point C.  Let us now consider the point A. As we enter the phase diagram switching
on the gauge-Higgs interaction, we observe that the first order transition is still
present and that the critical coupling slowly shifts towards smaller values. In figure
\ref{critG2HYM}a we show the probability distribution for the Polyakov loop at 3 values
of the gauge coupling -- increasing values from bottom to top -- at fixed gauge-Higgs
coupling $\kappa=1.3$. Two coexisting phases show up and their relative relevance switches
as the temperature increases: this gives a clear signal of a first order transition.
\vskip-.5cm

However, as we consider larger values of $\kappa$, the first order transition weakens and
then the phase transition is washed out to a crossover. For instance, in figure
\ref{critG2HYM}b, we show the probability distributions of the Polyakov loop at 3
different values of the gauge coupling (increasing values from bottom to top) at $\kappa=1.5$.
The distributions are quite broad and we do not find any indication of a phase
transition. However, we can not exclude that, performing numerical simulations on much
larger volumes, a weak first order transition could show up. 

If we further increase the parameter $\kappa$, the phase transition starts to show up
again and it stays there up to $\kappa=\infty$, where the limit of the $SU(3)$
Yang-Mills theory is attained. Note that the probability distribution of the real part of
the Polyakov loop at the deconfinement phase transition in $SU(3)$ Yang-Mills theory is
characterised by 3 peaks: one corresponding to the confined phase and the other two
associated with the deconfined one. One of the peaks of the deconfined phase is related to
the trivial $\Z(3)$ center element while the other one is the projection on the real axis
of the two non-trivial $\Z(3)$ center elements.  We expect that a similar 3-peak structure
should also progressively show up in the $G(2)$ gauge-Higgs system when the parameter
$\kappa$ is large enough. However, it is important to point out that, as long as $\kappa$
has finite values, the peak corresponding to the non-trivial $\Z(3)$ center elements is
metastable and there is a free energy difference proportional to the volume with respect
to the other deconfined peak. Hence, the second deconfined peak plays no role in the
critical behaviour since, in the thermodynamic limit, it would be absent. Nevertheless, in
a finite volume, as it is the case when performing numerical simulations, the second
deconfined, metastable peak should be observable for sufficiently large $\kappa$. Indeed, as we
can see in figure \ref{critG2HYM}c, the 3 peak structure shows up. Similar to the two
previous plots, we report the probability distribution for the Polyakov loop at 3
different values of the gauge coupling, at fixed $\kappa=4.0$.

\section{Conclusions}
Understanding the phenomenon of confinement and the features of the deconfinement phase
transition is an open, challenging problem. In this paper we have discussed the role
played by the center of the gauge group: we have approached the problem from the most
general perspective of considering all possible choices for the gauge symmetry group.
Then we have focused our attention on the case of $G(2)$. The group $G(2)$ is the
smallest, simply-connected group with a trivial center and $G(2)$ Yang-Mills theory is the
simplest pure gauge theory with no non-trivial 't~Hooft flux vortices. We have performed
Monte Carlo simulations in $G(2)$ Yang-Mills theory in (3+1) dimensions, finding numerical
evidence for a first order deconfinement phase transition. This result is consistent with
our conjecture that the size of the gauge group and not the center plays a relevant role
in determining the order of the deconfinement transition.

By exploiting the Higgs mechanism, one can study the connection between the deconfinement
phase transition of $G(2)$ Yang-Mills theory and the more familiar one of $SU(3)$
Yang-Mills theory. The numerical results indicate that the first order deconfinement
transition of $G(2)$ Yang-Mills theory weakens as the interaction with the Higgs field is
switched on. This supports our conjecture on the relation between the size of the group
and the order of the deconfinement transition. In fact the Higgs mechanism removes
progressively a number of degrees of freedom from the dynamics.

As the gauge-Higgs interaction is furtherly increased, a first order transition appears
again and it stays there until the $SU(3)$ limit is reached. We would interpret this
phase transition as an effect of the lack of a universality class with a $\Z (3)$ symmetry
in 3 dimensions and not as a results of the mismatch of the number of the relevant degrees
of freedom in $SU(3)$ Yang-Mills theory between the confined and the deconfined phases.

\emph{Acknowledgements} The study presented in this paper has been carried out in
collaboration with Uwe-Jens Wiese: an extended and detailed report will appear soon. I
gratefully acknowledge Philippe de Forcrand for many discussions and suggestions; I would
also like to thank Peter Minkowski for his challenging questions and his continuous
interest in the results of this project. Useful discussions with Kieran Holland are
acknowledged too.

\end{document}